\documentclass[reprint,aps,prb,superscriptaddress]{revtex4-1}

\usepackage{graphicx} 
\usepackage{dcolumn}
\usepackage{bm}
\usepackage{amssymb,amsmath}
\usepackage{epstopdf}

\begin{document}

\title{Revealing Optically Induced Magnetization in SrTiO$_3$ using Optically Coupled
SQUID Magnetometry and Magnetic Circular Dichroism}

\normalsize
\author{W.~D.~Rice}
\affiliation{National High Magnetic Field Laboratory, Los Alamos
National Laboratory, Los Alamos, NM 87545, USA}
\author{P.~Ambwani}
\affiliation{Department of Chemical Engineering and Materials Science,
University of Minnesota, Minneapolis, MN 55455, USA}

\author{J.~D.~Thompson}
\affiliation{Materials Physics and Applications Division, Los Alamos
National Laboratory, Los Alamos, NM 87545, USA}
\author{C.~Leighton}
\affiliation{Department of Chemical Engineering and Materials Science,
University of Minnesota, Minneapolis, MN 55455, USA}
\author{S.~A.~Crooker}
\affiliation{National High Magnetic Field Laboratory, Los Alamos
National Laboratory, Los Alamos, NM 87545, USA}
\date{\today}

\begin{abstract}
In this work, we study the time- and temperature-dependence of
optically induced magnetization in bulk crystals of slightly oxygen-deficient
SrTiO$_{3-\delta}$ using an optically coupled SQUID magnetometer. We
find that a weak ($\sim$5$\times$10$^{-7}$~emu) but extremely
long-lived (hours) magnetic moment can be induced in
SrTiO$_{3-\delta}$ at zero magnetic field by circular-polarized
sub-bandgap light.  We utilize this effect to demonstrate that
SrTiO$_{3-\delta}$ crystals can be used as an optically addressable
magnetic memory by writing and subsequently reading magnetic
patterns with light. The induced magnetization is
consistent with that of a polarized ensemble of independent oxygen-vacancy-related complexes, rather than from
collective or long-range magnetic order.
\end{abstract}


\maketitle


The developing field of complex oxide electronics is intimately tied
to its most widely studied member, strontium titanate
(SrTiO$_3$)~\cite{RameshMRSBulletin2008, MannhartScience2010,
ChambersAdvMat2010, ZubkoAnnRevCondPhys2011, HwangNatureMat2012}.
While the electronic and optical properties of bulk SrTiO$_3$ have
been thoroughly investigated~\cite{MullerPRB1979, LeePRB1975,
FaughnanPRB1971, WildPRB1973}, SrTiO$_3$ has recently drawn renewed
research interest mainly driven by the discovery of magnetism and
superconductivity at epitaxial interfaces between SrTiO$_3$ and
other complex oxides~\cite{HwangNatureMat2012,
BrinkmanNatureMat2007, DikinPRL2011, LiNaturePhys2011,
BertNaturePhys2011, AriandoNatureComm2011, MoetakefPRX2012,
LeeNatureMat2013}.  Because oxygen vacancies, $V_{\rm O}$, are
easily formed in SrTiO$_3$ and act as electron donors, it has been
widely conjectured that $V_{\rm O}$ may play a key role in these
emergent electronic and magnetic phenomena~\cite{MullerNature2004,
EcksteinNatureMat2007, KalabukhovPRB2007, ShenPRB2012,
PavlenkoPRB2012}. Interest in nominally non-magnetic bulk SrTiO$_3$
has been further fueled by observations of the Kondo effect and
magnetization in ionically gated SrTiO$_3$
crystals~\cite{LeePRL2011}.

Recently we reported~\cite{RiceNatureMater2014} the observation of
persistent optically induced magnetization in bulk crystals of
slightly oxygen-deficient SrTiO$_{3-\delta}$. Specifically,
circularly polarized light at sub-bandgap wavelengths between 400~nm
and 500~nm was found to induce a magnetic moment (in zero magnetic
field) that persisted from seconds at $\sim$18~K to several hours
below 10~K. This magnetization was attributed to a partial spin
polarization within the ground state of local $V_{\rm O}$-related
complexes. While these effects were investigated primarily using
magnetic circular dichroism (MCD) spectroscopy, critical time- and
temperature-dependent behaviors were directly confirmed using a
variant of conventional SQUID
magnetometry~\cite{RiceNatureMater2014}. Moreover, it was
demonstrated that detailed spatial magnetic patterns could be
written into and read from SrTiO$_{3-\delta}$ using light alone.

The primary intent of this paper is therefore to provide details of
the optically coupled SQUID magnetometry technique that was utilized
to confirm the presence of optically induced magnetization in our
SrTiO$_{3-\delta}$ crystals, and also to fully describe the
MCD-based optical system that was used to write and read magnetic
patterns in SrTiO$_{3-\delta}$. Detailed experimental schematics are
shown, along with data that further support a
scenario in which the induced magnetic moment arises \emph{not} from
long-range or collective magnetic order, but rather (and more
simply) from a long-lived spin polarization within an ensemble of
localized and independent $V_{\rm O}$-related complexes.

In these studies, nominally pure 500~$\mu$m thick SrTiO$_3$ (100)
crystals from MTI Corporation were annealed in ultra-high vacuum
(oxygen partial pressure $< 10^{-9}$~Torr) at temperatures between
650-750$^{\circ}$C, conditions that promote diffusion of oxygen out of the
lattice~\cite{SpinelliPRB2010}. This annealing (\emph{i.e.},
reduction) treatment was used to make a set of nine slightly
oxygen-deficient SrTiO$_{3-\delta}$ crystals with varying $V_{\rm
O}$ densities. To measure the $V_{\rm O}$ density, we soldered
indium contacts to the corners of each sample in a van der Pauw
geometry and measured the electron concentration, $n$, using
longitudinal resistivity and/or Hall-effect transport studies.  Assuming that
every $V_{\rm O}$ contributes one to two electrons to the
conduction band (in the simplest models), the approximate $V_{\rm O}$
concentration can be inferred from $n$.  In this work, $n$ ranged
from $\sim$10$^{14}$~cm$^{-3}$ to $8 \times 10^{17}$~cm$^{-3}$.

In our previous work~\cite{RiceNatureMater2014}, optically induced magnetization
in these SrTiO$_{3-\delta}$ crystals was studied primarily via the
technique of MCD spectroscopy. While non-zero MCD signals typically
imply the presence of time-reversal breaking (\emph{e.g.},
magnetism) \cite{StephensJChemPhys1970}, we felt that it was
important to independently confirm magnetism in our
SrTiO$_{3-\delta}$ crystals using a \emph{direct} SQUID-based probe
of the optically induced magnetic moment. Besides providing a
quantitative measure of the induced moment, SQUID magnetometry also
avoids certain artifacts (such as material gyrotropism) that can, in
certain circumstances, generate MCD or Kerr-effect signals that are
not related to real magnetism \cite{HosurPRB2013}.

To this end, we sought to perform SQUID magnetometry on our
SrTiO$_{3-\delta}$ crystals while illuminating the crystals with
light whose \emph{optical polarization} could be established and
controlled -- \emph{in situ} -- with a high degree of fidelity and
precision.  While many commercial SQUID systems provide options
allowing samples to be illuminated, this is typically achieved via
multi-mode optical fibers that scramble the polarization of the
transmitted light.  Therefore we constructed a portable optical
setup based on polarization control within a standard single-mode
fiber (SMF). Although bends and strain in a SMF generate
birefringence that can alter the polarization state of light
traveling within the SMF, the output polarization of the light is
not irretrievably scrambled. Rather, it has a definite and
measurable relationship to the input polarization, because the light
travels only along a single optical mode. Importantly, any
undesired birefringence or polarization changes in the SMF can be compensated
and ``undone" by carefully and intentionally straining the SMF using
a manual fiber polarization controller. Related fiber-based
approaches to couple polarized light to SQUID magnetometers have
been used in the past to study, for example, magnetic polarons in
diluted magnetic semiconductors ~\cite{AwschalomPRL1987}.

\begin{figure} [tbp]
\centering
\includegraphics [width = 3.2 in] {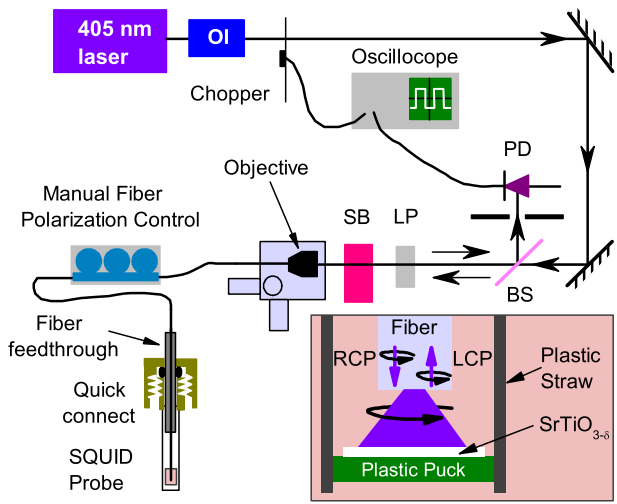}
\caption{Optically coupled SQUID magnetometry. Mechanically chopped
405~nm light from a laser diode is passed through an optical
isolator (OI) and is circularly polarized by a linear polarizer (LP)
and Soleil-Babinet compensator (SB) before being coupled into a
standard UV single-mode fiber. The light that is back-reflected from
the other end of the fiber (in the SQUID magnetometer) follows a
time-reversed path back through the fiber, SB, and LP and is
directed to a photodiode (PD) using a beamsplitter (BS). Its
intensity is zero (it is completely nulled by LP) if the optical
polarization at the \emph{end} of the fiber is exactly LCP or RCP. A
manual fiber polarization controller allows to compensate for
unwanted strain-induced birefringence in the fiber in order to
maintain circular polarization at the sample.  Inset: Detailed view
of fiber end near sample. The sample is mounted on a plastic Kel-F
(polychlorotrifluoroethylene) puck that is held by friction in a
plastic straw.} \label{Figure1}
\end{figure}

Figure~\ref{Figure1} depicts the optical setup that we used in
conjunction with a commercial Quantum Design MPMS SQUID
magnetometer. The system uses a standard probe whose top was
modified to admit a single-mode optical fiber (here, the bare fiber was
stripped of its plastic jacket and epoxied into a 10~cm length of brass tubing, that was
in turn fed through a standard vacuum quick-connect). The fiber emanating from the probe to the optical setup was carefully suspended far above the fiber feedthrough to minimize any strain or twisting of the fiber when the SQUID probe translates vertically during measurements; in doing so we find that the optical polarization is unaffected. Optical polarization conditioning and monitoring was achieved with
components on a portable 300~mm $\times$ 600~mm optical breadboard that was
located near the SQUID.  Circularly polarized 405~nm laser light was
coupled into a standard UV bare SMF using an aspheric objective lens
and a fiber launcher with piezoelectric actuators.  Free-space
circular polarization was produced before the fiber using a linear
polarizer (LP) and a Soleil-Babinet compensator (SB) on a rotational
mount. To compensate for unwanted birefringence in the SMF due to
bending and strain, we threaded the fiber through a three-paddle
manual fiber polarizer controller.

To infer the polarization state of the light at the sample, we monitored the light that
was back-reflected from the other end of the fiber (in the SQUID, just above the sample).  This back-reflected light
travels a time-reversed path back through the optical system (fiber,
SB, and LP), where it is picked off using a beamsplitter and
measured with a photodiode (PD) and oscilloscope.  Mechanically
chopping the beam provides a convenient baseline determination on
the oscilloscope trace.

The crucial point is that the back-reflected light will be polarized
exactly orthogonal to the linear polarizer LP (giving zero intensity
at the photodiode) \emph{only} when the light at the end of the
fiber -- and therefore at the sample -- is circularly polarized
(because only circular polarized light exactly reverses its helicity
upon reflection). Thus, manually straining the fiber to compensate
for any unwanted birefringence in the SMF and zeroing the
back-reflected intensity at the photodiode guarantees circularly
polarized light at the sample. Once one sense of circularly
polarized light is obtained, rotating the Soleil-Babinet compensator
by 45 or 90 degrees produces linear or oppositely circularly
polarized light at the sample, respectively (and a corresponding
maximum and second minimum of the back-reflected intensity). As might be expected, any significant movement or bending of the fiber during an
experiment requires a re-adjustment and balancing of the manual
polarization controller. This setup therefore utilizes the
back-reflected intensity as a continuous and \emph{in situ} monitor
of the optical polarization at the sample in the SQUID.

\begin{figure} [tbp]
\centering
\includegraphics [width = 2.7 in] {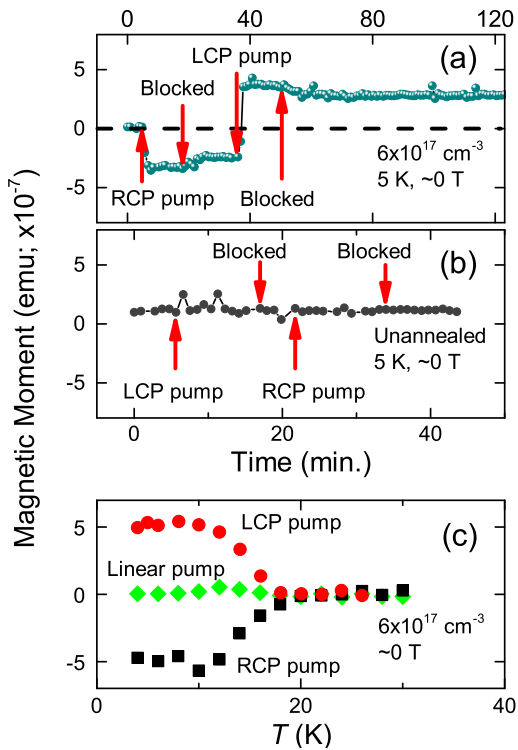} 
\caption{(a) The temporal evolution of the optically induced
magnetic moment in SrTiO$_{3-\delta}$, as measured by the optically
coupled SQUID, shows that magnetization at 5~K is maintained for
hours even after the light has been blocked.  RCP and LCP light
induce equal, but opposite, magnetic moments.  (b) No optically
induced magnetic moment is observed when the $V_{\rm O}$ density is
very low, as in unannealed SrTiO$_3$.  (c) Temperature-dependent
optically induced magnetic moment for RCP (black), LCP (red), and
linearly (green) polarized pumping as measured by SQUID magnetometry
(adapted from Rice \emph{et al.}~\cite{RiceNatureMater2014}, with permission from Nature Publishing Group).}
\label{Figure2}
\end{figure}

We used the optically coupled SQUID magnetometer to perform both
time- and temperature-dependent magnetometry on oxygen-deficient
SrTiO$_{3-\delta}$ crystals and also on unannealed (as-received)
SrTiO$_3$. Typical sample sizes were approximately
3~mm$\times$3~mm$\times$0.5~mm with measurements of the magnetic
moment being collected roughly every 90~seconds in zero magnetic
field with an average illumination power of $\sim$200~$\mu$W.
Figure~\ref{Figure2}(a) displays how the magnetic moment evolves
during and after pumping with circularly polarized light. For both
RCP and LCP light, the magnitude of the optically induced magnetic
moment is $\sim$$5\times$10$^{-7}$~emu, with RCP and LCP light
creating equal but oppositely-oriented magnetic moments. At low
temperatures, the magnetization persists for over an hour even after
the light is blocked, in agreement with MCD
results~\cite{RiceNatureMater2014}. To ensure that only samples with
oxygen vacancies produced this effect, we also tested an unannealed
(as-received) SrTiO$_3$ crystal under the same conditions. As seen
in Fig.~\ref{Figure2}(b), no magnetic moment is induced regardless
of the polarization of light.  The slight offset in the measured
magnetic moment is likely due to the small remnant magnetic field of
the superconducting magnet ($\sim$10$^{-4}$~T) and the intrinsic
diamagnetism of the sample and the plastic puck on which it was
mounted.

Temperature-dependent magnetization for RCP, LCP, and
linearly polarized illumination was also investigated.  As shown in
Fig.~\ref{Figure2}(c), the optically induced magnetic moment $M\left(T\right)$ undergoes a steep increase
below 18~K when continuously pumped with either RCP or LCP light;
however, no net magnetization is observed for linearly polarized
illumination.  Again, RCP and LCP light induce equal and opposite
magnetic moments.  $M$ saturates when $T$ decreases below
$\sim$10~K, a behavior that agrees well with MCD
studies~\cite{RiceNatureMater2014}, and which is shown below to be
independent of the $V_{\rm O}$ density.

\begin{figure} [tbp]
\centering
\includegraphics [width = 2.7 in] {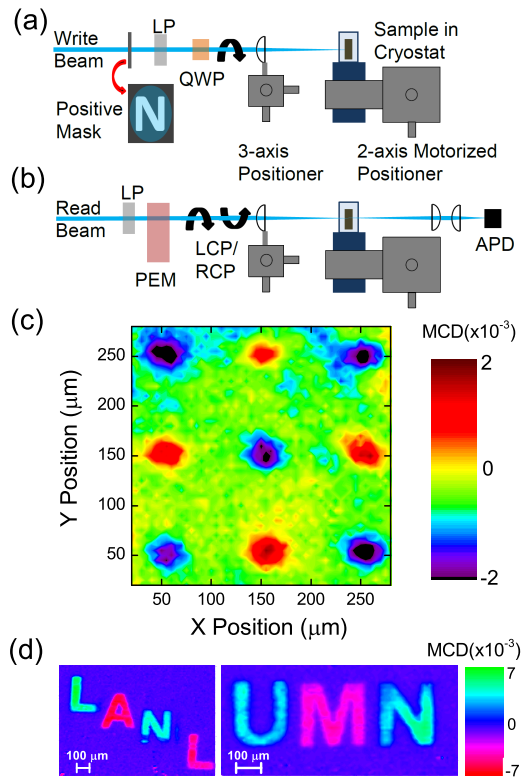}
\caption{(a) Experimental configuration used to optically write
magnetic images in SrTiO$_{3-\delta}$ at zero field.  A linear
polarizer (LP) and quarter-wave plate (QWP) are used to create RCP
or LCP light that can be imaged onto the sample through a positive
mask. (b) Setup used to optically read magnetic images using
raster-scanned MCD.  A photo-elastic modulator (PEM) produces
alternating RCP/LCP light that is transmitted through the sample and
detected by an avalanche photodiode (APD). (c,d) 2-D images of
detected magnetic patterns in SrTiO$_{3-\delta}$; here, the
magnetization is reversed at each subsequent dot/letter (Fig.~3d adapted from Rice \emph{et al.}~\cite{RiceNatureMater2014}, with permission from Nature Publishing Group).}
\label{Figure3}
\end{figure}

Optically producing a long-lived magnetic moment in zero applied
magnetic field is a potentially exciting development both
scientifically and technologically.  Figure~\ref{Figure3} demonstrates that
SrTiO$_{3-\delta}$ can potentially be utilized as a component in an
optically-addressable magnetic memory device. We created an imaging
system that can optically write and read spatial magnetic patterns
in SrTiO$_{3-\delta}$ [Fig.~\ref{Figure3}(a,b)]. For this
demonstration, we detected either simple  magnetic ``dots" or more complex
magnetic patterns that were written with 400~nm LCP/RCP light
[Figs.~\ref{Figure3}(c,d)]. Writing was typically accomplished by
passing RCP or LCP light through a positive mask and imaging it on
to a SrTiO$_{3-\delta}$ crystal at low temperature. To read these
magnetic patterns, a raster-scanned optical MCD probe was used to
measure magnetization as a function of position.

MCD spectroscopy is an all-optical technique that measures the
normalized difference between transmission (T) of RCP and LCP light:
$\rm{MCD} \propto \frac{T_{\rm RCP} - T_{\rm LCP}}{T_{\rm RCP} +
T_{\rm LCP}}$. Non-zero MCD signals are usually attributed to broken
time-reversal symmetry (\emph{e.g.},~magnetism) and can be used to
study a wide variety of magneto-optical
phenomena~\cite{StephensJChemPhys1970}. A key benefit of MCD methods
over conventional (SQUID, for example) magnetometry is its detailed
spectral dependence, which provides information on the energies of polarizable and magneto-optically active states. As previously shown~\cite{RiceNatureMater2014}, optically induced magnetization in
SrTiO$_{3-\delta}$ is most clearly revealed in the wavelength range
between 400-500~nm (just below the band edge), with a particularly
strong response at 425~nm. Therefore in the MCD images shown here,
probe light at 425~nm was modulated between RCP and LCP at 50~kHz by
a photo-elastic modulator (PEM) and was mechanically chopped at 137
Hz to facilitate lock-in detection of T$_{\rm RCP} - $T$_{\rm LCP}$
and T$_{\rm RCP} + $T$_{\rm LCP}$, respectively. This probe light was
focused and raster-scanned across the sample to construct a
two-dimensional image of magnetization. In these experiments, the
minimum size of the magnetic patterns was effectively limited by the
large thickness of the samples; we selected optics for which the
Rayleigh range of focused pump and probe light was commensurate with
the 500 $\mu$m thickness of the SrTiO$_{3-\delta}$ crystals. These
magnetic images can be erased by heating the SrTiO$_{3-\delta}$
crystal above $\sim$20~K.

We now present three pieces of new data that further elucidate the
underlying nature of the optically induced magnetization in
SrTiO$_{3-\delta}$, and which further support a scenario in which
the magnetization arises \emph{not} from collective or long-range
interactions, but rather (and more simply) from a metastable spin
polarization in the ground states of an ensemble of independent $V_{\rm O}$-related complexes. We note that $V_{\rm O}$ may be forming defect complexes with residual elemental impurities (for example, Fe) that are present at the tens of parts-per-million level even in nominally pure commercial SrTiO$_3$ crystals. A detailed analysis of trace impurities in our crystals was presented in our recent work\cite{RiceNatureMater2014}.

Figure 4(a) shows optically induced magnetization as a function of temperature for
three different SrTiO$_{3-\delta}$ crystals having $V_{\rm O}$
concentrations that vary over three orders of magnitude. Whether the
$V_{\rm O}$ density is large ($6 \times 10^{17}$ cm$^{-3}$) or small ($2\times
10^{14}$ cm$^{-3}$), the optically induced magnetization appears at the
\emph{same} temperature of about 18~K.  If the magnetic moment were
due to long-range interactions  between $V_{\rm O}$ complexes
(for instance, due to itinerant ferromagnetism), then some difference in
this critical temperature would be expected as the $V_{\rm O}$  density (and therefore also the electron density $n$) changes, as was
shown for the ferromagnetic semiconductors PbSnMnTe,
Ge$_{1-x}$Mn$_x$Te, or Ga$_{1-x}$Mn$_x$As when the density of
itinerant carriers was varied ~\cite{StoryPRL1986, FukumaAPL2002,
KoshiharaPRL1997}. In contrast, the data in Figure 4(a) are consistent
with a single-entity effect, where the polarization of individual
$V_{\rm O}$-related complexes becomes extremely long-lived below
18~K.

\begin{figure} [tbp]
\centering
\includegraphics [width = 2.5 in] {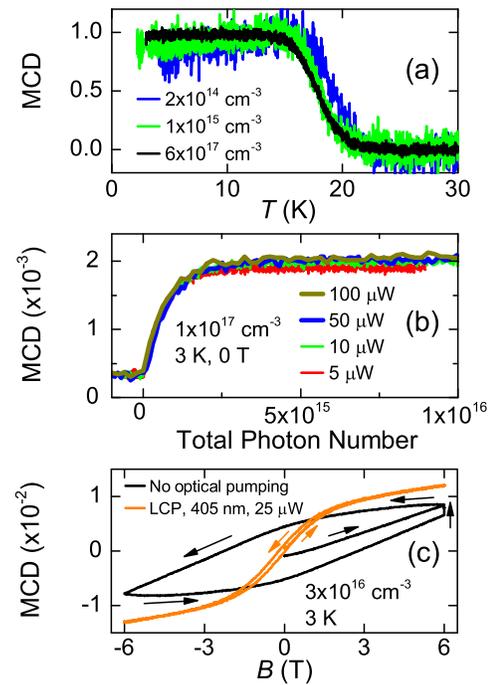}
\caption{(a) Temperature-dependent magnetization (as measured by MCD
at 425~nm) under continuous pumping with 405~nm LCP light for
SrTiO$_{3-\delta}$ crystals having different $V_{\rm O}$ densities. The
curves are normalized in magnitude for comparison. Regardless of the
$V_{\rm O}$ concentration, the form of $M(T)$ is the
same. (b) Optically induced magnetization in SrTiO$_{3-\delta}$ as a
function of total photon number for different illumination
intensities. (c) Magnetization, as measured by MCD at 425~nm, versus
magnetic field with (orange curve) and without (black curve) LCP
optical pumping. Hysteresis is due to slow magnetization dynamics;
faster equilibration occurs when the  system is continuously pumped
with circularly polarized light.} \label{Figure4}
\end{figure}

Figure 4(b) shows that the build-up and saturation of
optically induced magnetization in SrTiO$_{3-\delta}$ depends only
on the total number of photons incident on the sample (at a given
wavelength), rather than explicitly on the duration or intensity of
illumination. In this data we monitored the temporal build-up of
magnetization following illumination with 405~nm light having
intensity ranging from 5~$\mu$W to 100~$\mu$W. Plotting the induced
moment as a function of the total number of photons received (rather
than as a function of time) collapses all data traces on to a single
curve. Again, these data are consistent with a non-interacting
ensemble of $V_{\rm O}$-related complexes that, with some finite
cross-section, can become spin-polarized by circularly polarized light.
Moreover, these data show that any changes in the background electron density $n$ due to illumination has little
effect.

Finally, Fig.~\ref{Figure4}(c) shows the measured MCD signal from a
SrTiO$_{3-\delta}$ crystal as a function of applied magnetic field
$B$ in the Faraday geometry, for the two cases of continuous optical LCP
pumping and no optical pumping. Although it is
tempting to associate the observed hysteresis with long-range
magnetic order, the observed open hysteresis loops are due to the slow
magnetization dynamics that were identified in, for example, Fig.~\ref{Figure2}. In the absence of optical pumping, the magnetization requires
hours to relax and re-equilibrate.  With optical pumping, the
characteristic timescales are much faster, of order seconds to
minutes depending on the illumination intensity. In this latter case,
the sample magnetization is better able to remain in approximate
equilibrium as the magnetic field is swept at 1 tesla/minute, and
the observed hysteresis loop is narrow.

In summary, we developed and used an optically coupled SQUID
magnetometer to deliver and monitor circularly polarized light to an
\emph{in situ} sample.  This setup was used to investigate the
temperature- and time-dependent magnetization in oxygen-deficient
SrTiO$_{3-\delta}$.  All data strongly suggest that the optically
induced magnetization arises from a long-lived spin polarization in
an ensemble of localized and independent $V_{\rm O}$-related
complexes, rather than from collective effects such as
ferromagnetism.  We demonstrated the technological promise of
persistent optically induced magnetization  by optically writing,
storing, and optically reading magnetization in a SrTiO$_{3-\delta}$
crystal in zero magnetic field, pointing to exciting new frontiers
in complex oxide electronic and magneto-optical devices.
\newline
\newline
\noindent
\textbf{Acknowledgements}
\newline
This work was supported by the Los Alamos LDRD program under the
auspices of the US DOE, Office of Basic Energy Sciences, Division of
Materials Sciences and Engineering. Work at UMN supported in part by
NSF under DMR-0804432 and in part by the MRSEC Program of the NSF
under DMR-0819885.

\end{document}